\documentclass[5p,sort&compress,times,fleqn]{elsarticle}

\usepackage{amsmath,amssymb}
\usepackage[breaklinks]{hyperref}
\usepackage[english]{babel}
\usepackage[utf8]{inputenc}
\usepackage{graphicx}
\usepackage{color}
\usepackage{amsfonts,amsthm}
\usepackage{bm,bbm}
\usepackage{ulem}

\renewcommand{\emph}{\textit}
\newcommand{\be}{\begin{equation}}
\newcommand{\ee}{\end{equation}}
\newcommand{\ba}{\begin{eqnarray}}
\newcommand{\ea}{\end{eqnarray}}
\def\bs{\begin{subequations}}
\def\es{\end{subequations}}

\def\a{\alpha}

\def\de{\delta}

\def\la{\lambda}

\def\om{\omega}

\def\s{\sigma}

\def\cP{\mathcal{P}}

\def\cV{\mathcal{V}}

\def\ds{d_{\rm S}}
\def\dh{d_{\rm H}}

\def\p{\partial}

\newcommand{\Eq}[1]{(\ref{#1})}
\def\com{\color{magenta}}
\def\cob{\color{blue}}
\newcommand{\book}[5]{{#1}, #2, #3, #4, #5}

\newcommand{\oarX}[1]{\href{http://arxiv.org/abs/#1}{{\ttfamily\com arXiv:#1}}}
\newcommand{\arX}[1]{\href{http://arxiv.org/abs/#1}{{\ttfamily\com arXiv:#1}}}
\newcommand{\doin}[6]{\href{http://dx.doi.org/#1}{{\cob #2 #3 {#4} (#6) #5}}}
\newcommand{\doinn}[5]{\href{http://dx.doi.org/#1}{{\cob #2 {#3} (#5) #4}}}
\newcommand{\doij}[5]{\href{http://dx.doi.org/#1}{{\cob #2 #3 (#5) #4}}}

\newcommand{\procsinm}[5]{in \emph{#1}, Ed.\ by #2, #3, #4, #5}
\newcommand{\tia}[1]{#1,}

\def\lp{\ell_{\rm Pl}}

\def\rmd{d}

\journal{Physics Letters B}

\begin{document}

\title{Imprint of quantum gravity in the dimension and fabric of spacetime}

\author{Giovanni Amelino-Camelia}
\ead{giovanni.amelino-camelia@roma1.infn.it}
\address{Dipartimento di Fisica, Universit\`a di Roma ``La Sapienza", P.le A. Moro 2, 00185 Roma, Italy}
\address{INFN, Sez.~Roma1, P.le A. Moro 2, 00185 Roma, Italy}

\author{Gianluca Calcagni}
\ead{calcagni@iem.cfmac.csic.es}
\address{Instituto de Estructura de la Materia, CSIC, Serrano 121, 28006 Madrid, Spain}

\author{Michele Ronco}
\ead{michele.ronco@roma1.infn.it}
\address{Dipartimento di Fisica, Universit\`a di Roma ``La Sapienza", P.le A. Moro 2, 00185 Roma, Italy}
\address{INFN, Sez.~Roma1, P.le A. Moro 2, 00185 Roma, Italy}

\begin{abstract}
We here conjecture that two much-studied aspects of quantum gravity, \textit{dimensional flow} and \textit{spacetime fuzziness}, might be deeply connected. We illustrate the mechanism, providing first evidence in support of our conjecture, by working within the framework of multifractional theories, whose key assumption is an anomalous scaling of the spacetime dimension in the ultraviolet and a slow change of the dimension in the infrared. This sole ingredient is enough to produce a scale-dependent deformation of the integration measure with also a fuzzy spacetime structure. We also compare the multifractional correction to lengths with the types of Planckian uncertainty for distance and time measurements that was reported in studies combining quantum mechanics and general relativity heuristically. This allows us to fix two free parameters of the theory and leads, in one of the scenarios we contemplate, to a value of the ultraviolet dimension which had already found support in other quantum-gravity analyses. We also formalize a picture such that fuzziness originates from a fundamental discrete scale invariance at short scales and corresponds to a stochastic spacetime geometry.
\end{abstract}

\date{June 21, 2017}


\maketitle

\section{Introduction and main goal} 

The landscape of quantum gravity (QG) looks like a variegated compound of approaches that start from different conceptual premises and use different mathematical formalisms (see, {e.g.}, Refs.\ \cite{Ori09,Fousp,Zwi09,rov07,thi01,Per13,GiSi,AmMa,AGJL4,Dow13,LaR5,NiR,RSnax,ADKLW,BIMM,Hor3,CES,Tom97,Mod1,BGKM,CaMo2}). Rather surprisingly, despite this heterogeneity, over the past few years  a generic prediction has emerged: \textit{dimensional flow} \cite{tH93,Car09,MoN,fra1,CaMo1,COT2,COT3,trtls,dimLQG1,dimLQG,Benedetti,spectDim,dimMDR,padma1,astrid,Co1,BBMM,Co2}, i.e., a change of spacetime dimension with the scale of the observer. In all QG models, the dimensionality of spacetime exhibits a dependence on the scale, changing (or ``flowing'') from the topological dimension $D$ in the infrared (IR) to a different value in the ultraviolet (UV). So far, there has been no deep explanation for this universal property. Understanding its origin is just as important as looking for its physical characterization, needed to relate the flow of dimensions to physical observables.

We here put forward and motivate the conjecture that dimensional flow is directly related to the presence of limitations on the measurability of distances close to the Planck length $\lp = \sqrt{G\hbar/c^3}$, a feature (\textit{spacetime fuzziness}) which has been of interest for QG research for decades \cite{ven,padma,kon,ellis,gara,alu,ngdam,amelino}. More precisely, we shall provide preliminary ``theoretical evidence'' in support of a connection between the number of spacetime dimensions in the UV and the form of the uncertainty on spacetime distances. Important from our perspective is the fact that such a connection might set the stage for a role for dimensional flow in QG phenomenology \cite{gacLRR}. Indeed, it has been shown that, in some cases, spacetime fuzziness could be investigated in ongoing and forthcoming experiments, even if the fuzziness is introduced at the Planck scale. This was first explored in analyses of the interferometers used for gravity-wave searches \cite{gw1,gw2,gacLRR}, and more recently is focusing mainly on the implications of fuzziness for the formation of halo structures in the images of distant quasars \cite{quasar,gacLRR}.

\section{Example: multifractional theories} 

We provide preliminary support for our conjecture within the context of {multifractional theories} \cite{fra1,first}  fully reviewed in \cite{NewRev}. These are a class of field theories of matter and gravity where spacetime is ``anomalous'' and changes properties with the probed scale, in a way similar to a multifractal. While in other quantum gravities dimensional flow is a derived property not required \emph{a priori}, here it is part of the definition of the framework. Thanks to their peculiar properties, these field theories living on a multifractal spacetime reproduce a wealth of phenomena found in QG. In particular, the running of dimensions is produced by an integration measure of the type $\rmd^Dq(x):=\rmd q^0(x^0) \rmd q^1(x^1)\cdots \rmd q^{D-1}(x^{D-1})= \p_0q^0\rmd x^0 \p_1 q^1\rmd x^1\cdots \\ \p_{D-1}q^{D-1}\rmd x^{D-1}$. The factorizable form is assumed for technical reasons \cite{NewRev} not especially important here, while the specific form of the distributions $q^\mu(x^\mu)$ is obtained by requiring that dimensional flow is slow at large scales. This assumption (spacetime dimension almost constant in the IR), true in all quantum gravities without known exception, is at the core of a result we will invoke often later, the second flow-equation theorem \cite{first} (a ``first'' version holds for nonfactorizable measures). An approximation of the full measure, which is physically nonrestrictive but will be refined later, is the binomial space-isotropic profile
\begin{equation}\label{multimeas}
q^\mu(x^\mu)\simeq (x^\mu-\overline{x}^\mu) + \frac{\ell_*}{\a_\mu}\left|\frac{x^\mu-\overline{x}^\mu}{\ell_*} \right|^{\a_\mu}\,,
\end{equation}
where the index $\mu$ is not summed over and takes values $0,1,2,\dots, \\ D-1$. For simplicity, we assume $\a_\mu=\delta_{0 \mu} \a_0 + (1-\delta_{0 \mu}) \a$, i.e., the exponents $\a_{\mu\neq 0}$ associated with spatial directions have all the same value $\a$; moreover, we also enforce $0<\a_0,\a<1$, to avoid negative dimensions and obtain the correct IR limit \cite{NewRev}. Note that \Eq{multimeas} is uniquely determined parametrically as soon as dimensional flow is switched on and is slow (almost constant spacetime dimension) in the IR \cite{first}. This means that different models of quantum gravity can predict different values of the parameters $\a_\mu$ and $\ell_*$ (plus other parameters that appear in the full expression at mesoscopic scales \cite{first}), but the general form of the measure as a parametric profile are the same and given by \Eq{multimeas}. The only ambiguity left undecided by the second flow-equation theorem is a shift in the coordinates, represented by the given point $\overline{x}^\mu$. This shift ambiguity is a puzzling aspect from the viewpoint of interpretation, since it is a sort of preferred point in the universe. However, our results will neutralize this feature and embed it into a more amenable physical interpretation. We will comment on this shortly.

Depending on the symmetries of the Lagrangian, there are four possible multifractional theories, classified according to the derivative operators appearing in kinetic terms. Here we will concentrate on two theories with the same asymptotic expression for lengths, with so-called $q$- and fractional derivatives. For the purposes of this paper, suffice it to say that $q$-derivatives are defined as $\p_{q^\mu}=(\rmd q^\mu/\rmd x^\mu)^{-1}\p_\mu$. Details on fractional derivatives are discussed in \cite{NewRev}.

To get the Hausdorff dimensions $\dh$ of spacetime, one computes the volume $\cV$ of a $D$-cube with size edge $\ell$, leading to the result that, if $\a_0=\a$ (as fixed by the arguments below), then $\cV=\int_{\rm cube} \rmd^Dq(x) \simeq \ell_*^D [({\ell}/{\ell_*})^D +({\ell}/{\ell_*})^{D\alpha}]$. Thus, we have $\dh \simeq D\alpha$ in the UV  ($\ell<\ell_*$). Here we have neglected mesoscopic contributions to $\cV$, which are not relevant to get the number of dimensions in the far UV \cite{frc2}. For the two multifractional theories considered here, it is not difficult to prove that, in the UV, the spectral dimension (the scaling of the return probability $\cP\sim\ell^{-\ds}$ measuring how likely it is to find a test particle in a neighborhood of its actual position when probing spacetime with an apparatus with resolution $1/\ell$) coincides with the Hausdorff dimension, $\ds\simeq D\a\simeq\dh$, for $\a_0=\a$ \cite{NewRev}. Both $\alpha$ and $\ell_*$ are free parameters of the theory with the only requirement that $\ell_*$ must be small enough to comply with experimental constraints \cite{NewRev}. As said above, the measure $q^\mu(x^\mu)$ is fixed by the second flow-equation theorem \cite{first}, but there remains an ambiguity related to the choice of a preferred frame, which amounts to the choice of $\overline{x}^\mu$ in Eq.\ \eqref{multimeas}. In fact, physical observables have to be compared in the picture with $x^\mu$ coordinates representing clocks and rods that do not adapt to the scale. This poses the so-called presentation problem \cite{trtls,NewRev}, which consists in the choice of the physical frame where Eq.\ \Eq{multimeas} is defined and observables are calculated.

\section{Connecting dimensional flow and fuzziness: first glimpse} 

As announced, we shall use multifractional theories as a testing ground for our conjecture. We shall seek a connection between dimensional flow in multifractional theories and the limitations on the measurability of spacetime distances obtained by many authors heuristically combining aspects of quantum mechanics (QM) and general relativity (GR) \cite{padma,gara,alu,ngdam,amelino}. It is noteworthy that the presence of these distance-measurement uncertainties, though originally discussed exclusively with heuristic reasoning, has  found confirmation in concrete QG theories in recent years (see, {e.g.}, Refs.\ \cite{Ori09,Fousp}), each of which realizes the corresponding UV features in very different ways \cite{tH93,Car09,fra1,NewRev}. The observations we here report can also be viewed as an explanation of why one gets a correct intuition about distance fuzziness even just resorting to the qualitative interplay of QM and GR. The link is provided by the fact that limitations on geometric measurements are intimately related to dimensional flow. As a byproduct of our analysis, we will also give a physical interpretation for the ambiguities of multifractional theories and select two sets of preferred values for $\a$ and $\ell_*$. Remarkably, in one of these cases, we obtain $\a = 1/2$ and, consequently, $\dh \simeq \ds \simeq 2$ in the UV, a value that has already been singled out for independent reasons in many QG studies (see Refs.\ \cite{LaR5,Hor3,CES,tH93,Car09,fra1,dimLQG1,spectDim,dimMDR,padma1,astrid} and references therein).

We focus on the $(1+1)$-dimensional theory with $q$-derivatives, a context where the analysis progresses more simply but without loss of any characteristic feature. Using Eq.\ \eqref{multimeas}, the reader can easily realize that the spatial distance between two points $A$ and $B$ is
\begin{equation}\label{dist}
L := \int_{x_{\rm A}}^{x_{\rm B}} \rmd q^1 = \ell+ \frac{1}{\alpha}\frac{\ell_*}{\ell} \left(\left | \frac{x_{\rm B}-\bar{x}}{\ell_*}  \right | ^\alpha -\left | \frac{x_{\rm A}-\bar{x}}{\ell_*}  \right | ^\alpha\right) ,
\end{equation}
with $\ell = x_{\rm B}-x_{\rm A}$. Thus, different presentations (i.e., different values of $\bar{x}$ \cite{NewRev,trtls}) give different results for the distance, although they do not change the anomalous scaling, which is solely governed by $\a$. Up to now, this has been regarded as a freedom of the model, but we here suggest that the presentation ambiguity should be viewed as a manifestation of spacetime fuzziness.

Four presentation choices have been identified as special among the others \cite{trtls}, but the second flow-equation theorem \cite{first} selects only two of these: the {initial-point presentation}, where $\bar{x} = x_{\rm A}$, and the {final-point presentation}, where $\bar{x} = x_{\rm B}$. In both cases, Eq.~\eqref{dist} simplifies in such a way that the difference between $L$ and the value $\ell$ that would be measured in an ordinary space is \cite{trtls}
\begin{equation}\label{multunc}
\de L_\a\simeq \pm\frac{\ell_*}{\a}\left(\frac{\ell}{\ell_*}\right)^\a\,,
\end{equation}
approximately valid in any space dimensions, where the plus sign is for the initial-point presentation and the minus is for the final-point presentation.

Strikingly, the  multifractional contribution to distances (\ref{multunc}) is of the same type of the lower bound on distances found by heuristically combining QM and GR arguments \cite{padma,gara,alu,ngdam,amelino}. In particular, in Ref.\ \cite{amelino}, one of us proposed an argument leading to a minimal length uncertainty $\de L\sim \sqrt{\lp^2 \ell/s}$, where $s$ is a length scale characterizing the measuring apparatus. Using a somewhat different line of reasoning, the authors of Ref. \cite{ngdam} suggested instead fluctuations of magnitude $\sim (\lp^2 \ell)^\frac{1}{3}$. Both of these well-studied scenarios for distance fuzziness match quantitatively the  multifractional contribution to distances (\ref{multunc}) upon adopting
\begin{equation}\label{a12}
\alpha = \frac{1}{2} \, , \qquad \ell_* = \frac{\lp^2}{s} \,,
\end{equation}
in agreement with Ref.\ \cite{amelino}, or
\begin{equation}\label{a13}
\alpha = \frac{1}{3} \, , \qquad \ell_* = \lp \,,
\end{equation}
in agreement with Ref.\ \cite{ngdam}.

This leads us to advocate a novel interpretation of \eqref{multunc}, such that it gives an intrinsic uncertainty on the measurement of spacetime distances. According to this interpretation, the initial-point presentation generates a positive fluctuation $+\delta L_\alpha$, while the final-point presentation produces a negative fluctuation $-\delta L_\alpha$, with the possibilities $\a = 1/2$ and $\a = 1/3$ being favored by the connection with \cite{ngdam,amelino} we are starting to build up.

The value \Eq{a12} has been already recognized as special for several theoretical reasons \cite{NewRev}. In particular, it gives the aforementioned result $\ds \simeq 2$ in the UV. What is more, the length scale $\ell_*$ turns out to be related to the Planck length. In the case \Eq{a12}, we have $\ell_*=\lp^2/s<\lp$, where $s$ is the observation scale. Thus, the dependence on the scales at which the measurement is being performed becomes explicit. This is exactly what is expected to happen in multifractal geometry and, in particular, in multifractional theories, where the results of measurements depend on the observation scale \cite{NewRev}. In the case \Eq{a13}, $\ell_*$ coincides with $\lp$. In both cases, the relation of $\ell_*$ with $\lp$ exposes the possibility of encoding highly nontrivial \emph{quantum} features within multifractional theories. A similar line of thought applies also to the time direction, which leads us to entertain the concrete possibility that the binomial measure should be isotropic in space and time, i.e., 
\be
\a_0=\a\,.
\ee

It is intriguing that, in the illustrative example for our main claim, a connection is established between a multifractional theory with a built-in dimensional flow (a feature usually derived, rather than assumed, in top-down approaches to QG) and uncertainties on distance measurements motivated by heuristic bottom-up approaches, combining just QM and GR principles without adding any hypothetical QG ingredient. We are thereby conjecturing that the connection between the form of dimensional flow and the form of spacetime fuzziness should have \emph{wider applicability}. However, also within the limits of our example some additional consistency checks are appropriate. A more in-depth analysis is needed in order to establish satisfactorily that, in multifractional models, both dimensional flow and spacetime fuzziness are obtained without introducing internal contradictions or external ingredients.

\section{Core of the connection: stochastic spacetime emerges}

From the multifractional perspective, the reinterpretation we are proposing is not arbitrary. In Ref.\ \cite{trtls}, it was observed that the theory with fractional derivatives describes spacetimes with a microscopic stochastic structure, i.e., a nowhere-differen\-tiable geometry where location of events (``points'' in space) cannot be determined with arbitrary accuracy and particle trajectories are nonsmooth. The presentation label $\bar x^\mu$ prescribes how integrals on stochastic spacetime variables can be per\-form\-ed, as in the It\^o--Stratonovich dilemma in random processes. Inspired by this, instead of defining as many physically inequivalent theories (but with the same anomalous scaling) as the number of presentations, and to choose one presentation among the others, one can take ``all presentations at the same time.'' In this case, the measures $\{q^\mu(x-\bar x^\mu)\,:\,\bar x^\mu \in\mathbbm{R}^D\}$ would not correspond to a class of (in)finitely many theories labeled by $\bar x^\mu$ all with the same anomalous scaling: they would be \emph{one} measure corresponding to \emph{one} theory with an intrinsic microscopic uncertainty. This \emph{stochastic view} holds only in the multifractional theory with fractional derivatives and also in the case with $q$-derivatives, which is an approximation of the former \cite{NewRev}. 

A direct and rigorous way to understand where stochasticity may come from in classical multifractional spacetimes is the following. A connection between a fractal and a stochastic structure has been advanced long ago by Nottale in his \emph{scale relativity} (e.g., Ref.\ \cite{Not08}). There, assuming that spacetime is fractal, the expression for a length was found to be $L=\ell+\zeta\ell_*(\ell/\ell_*)^\a$, where one discriminates between a deterministic differentiable part $\ell$ (the length on usual space) and a stochastic, nowhere-differentiable part $\zeta$. The latter is a wildly fluctuating random variable such that $\langle\zeta\rangle=0$ and $\langle\zeta^2\rangle=\mp 1$, depending on whether the distance is time- or spacelike. Since both scale relativity and multifractional spacetimes rely on a fractal geometry, it is not surprising that they lead to similar descriptions of  lengths. However, the original fractal-spacetime formulation of multifractional theories \cite{frc2} has been made much more solid thanks to a fundamental principle (slow IR dimensional flow) \cite{first} that reproduces the measure dictated by fractal geometry and fixes some of the free parameters of scale relativity. In particular, not only is the stochastic random variable $\zeta$ of Nottale's ``fractal'' length $L$ present in a more general multifractional length if we go beyond the approximation \Eq{multimeas} of a binomial measure, but it is also fixed by the second flow-equation theorem, in contrast with the \emph{ad hoc} variable $\zeta$ in scale relativity. In fact, considering the second-order truncation of the full measure determined by the flow-equation theorem \cite{first}, we have (index $\mu$ omitted everywhere)
\begin{equation}\label{qF}
q(x)=x+\frac{\ell_*}{\a}\left|\frac{x}{\ell_*}\right|^\a F_\om(x)\,,
\end{equation}
where $F_\om(x)=F_\om(\la_\om x)$ is a complex modulation factor encoding a fundamentally discrete spacetime symmetry $x\to \la_\om x$ in the far UV ($\la_\om$ is fixed). This symmetry arises as a consequence of the theorem, it is not imposed by hand, and Eq.\ \Eq{qF} is the generalization of \Eq{multimeas} (with $\bar x=0$ for simplicity; presentation does not affect the argument here) to higher orders in the flow equation \cite{first}. Requiring the measure to be real-valued, one has \cite{first,NewRev,frc2}
\bs\label{logos}\ba
\hspace{-.7cm}F_\om(x) &=&\sum_{n=0}^{+\infty}F_n(x),\qquad \om_n=\om n\,,\\
\hspace{-.7cm}F_n(x)&:=&A_n\cos\left(\om_n\ln\left|\frac{x}{\ell_\infty}\right|\right)+B_n\sin\left(\om_n\ln\left|\frac{x}{\ell_\infty}\right|\right)\!,
\ea\es
where $A_n$ and $B_n$ are constant amplitudes and $\lp\sim\ell_\infty\lesssim\ell_*$. The coordinate dilation of the discrete scale invariance is governed by the frequency $\om$, $\la_\om=\exp(-2\pi/\om)$. The log-oscillating structure is determined by the flow-equation theorem \cite{first}, while the simple but crucial linear relation $\om_n=\om n$ is determined by discrete scale invariance, the trade mark of iterative (also called deterministic) fractals \cite{frc2}. For phenomenological reasons, the modulation factor \Eq{logos} is usually approximated by only two frequencies, the zero mode $n=0$ [$F_0(x)=A_0={\rm const}$] and the $n=1$ mode. This approximation is quite effective in capturing the physical imprint of the logarithmic oscillations in particle-physics and cosmological observables \cite{NewRev}, but here we prefer to retain the full structure \Eq{logos}. Defining $y:=\ln|x/\ell_\infty|$ and taking the average $\langle f(y)\rangle:=(2\pi)^{-1}\int_0^{2\pi}\rmd y\,f(y)$, we get
\be
\langle F_\om\rangle=A_0\,,\qquad \langle F_\om^2\rangle=A_0^2+\sum_{n>0}\frac{A_n^2+B_n^2}{2}\,.
\ee
Thus, \emph{if we drop the zero mode} and set $A_0=0$, the profile $\tilde F_\om(x):=\sum_{n>0}F_n(x)$ reproduces Nottale's fractal coordinates upon the identification $\zeta=\tilde F_\om$. Since the sign and magnitude of the multiscale correction to lengths are modulated by log oscillations, the latter solve the presentation problem by making the presentation choice irrelevant. Moreover, for certain $n$-dependences of the amplitudes $A_n$ and $B_n$ (corresponding to introducing ergodic mixing phases in the oscillations), Eq.\ \Eq{logos} is a Weierstrass-type nowhere-differentiable function \cite{GlSo}. Non-differentiability is a key property of random distributions. As a result, we reach the neat conclusion that, in multifractional theories, the ``stochastic fluctuations'' of the geometry are provided by the logarithmic oscillatory modulation of the measure.

\section{Extending to other quantum gravities}

The logic so far has been to obtain fuzziness, an effect of quantum mechanics that can be found in quantum gravity, as a byproduct of the multifractal structure of spacetime. Assuming that spacetime has dimensional flow is sufficient to obtain some quantum-gravity effects tightly related with the quantization paradigm. In this respect, an intrinsic fractal structure of spacetime can, in some loose sense, ``replace'' quantum mechanics, inasmuch as it is responsible for uncertainty in time and distance measurements. In particular, this explains why classical multifractional theories can encode these features efficiently. An added advantage gained by this perspective is that, as ensured by the flow-equation theorem, the choice of such a fractal structure at short distances is unambiguous. However, by definition quantum mechanics cannot be dispensed with in quantum gravity, and fuzziness and dimensional flow are, at least in part, its consequences. Therefore, the consequence of the present results for quantum gravity is not a change in the main paradigm (gravity is still quantized, in each model by a different fashion) but, rather, the unification of two concepts, dimensional flow and fuzziness, so far considered as separate entities. 

In closing, let us offer some additional comments on how our observations might shed light on why the flow of dimensions in the UV is a universal property of QG approaches. Our findings indicate the possibility that dimensional flow is linked to distance fuzziness, whose form can be inferred from arguments combining QM and GR, without knowledge of the detailed features of one or another QG model. In this respect, spacetime fuzziness could be viewed in analogy with the Hawking temperature for black holes, also derived from semiquantitative model-independent arguments combining QM and GR.

Multifractional theories are particularly manageable for what concerns the structures that one needs to investigate in order to test our conjecture. Of course, this is the reason why we chose them as the example for this first exploratory study. The test may be harder in other formalisms of quantum gravity, but we hope that the encouraging results reported here will energize efforts in that direction. All the main elements of our arguments are already in place in some of the major proposals in the literature. In particular, string theory and nonlocal quantum gravity both realize dimensional flow \cite{Mod1,CaMo1} and coarse-grain (in the case of the string) or eliminate UV divergences (see \cite{MoRa} for nonlocal quantum gravity). Asymptotically-safe quantum gravity and the discrete-geometry, mutually related frameworks of loop quantum gravity, spin foams and group field theory all have dimensional flow \cite{AGJL4,NiR,RSnax,COT2,COT3,LaR5,dimLQG1,dimLQG} and implement fuzziness by the presence of minimal lengths, areas or resolutions \cite{rov07,RSc1,RSc2}. Maybe also causal dynamical triangulations \cite{AGJL4} realize fuzziness, as indicated by modified-dispersion-relation arguments \cite{Co2}. However, although a relation between anomalous dimensions and fuzzy features certainly seems to exist in these cases, as well as in general in QG \cite{padma1}, so far it has not been understood beyond the merely technical level. Revisiting those theories in search of a physical connection similar to that found here may help to clarify some of their formal aspects and even give new tools by which to extract testable phenomenology.

\section{Phenomenology} 

We conclude with the implications of our conjecture for phenomenology. If indeed our conjecture was confirmed, then the phenomenology would be empowered by the possibility of combining experimental bounds on dimensional flow and experimental bounds on fuzziness. For example, for multifractional theories the established bounds on dimensional flow \cite{NewRev} acquire the added significance of  bounds on the minimal resolution $1/\ell_*$ achievable. In turn, from Eq.\ \Eq{multunc} we can infer constraints on time-space isotropic $\ds$ (or $\dh$) using bounds on fuzziness \cite{gacLRR,gw1,gw2,quasar}. In fact, neglecting an $\mathcal{O}(1)$ numerical factor, Eq.\ \Eq{multunc} yields spacetime fuzziness of the form $\sigma \sim (\ell_*)^{1-\a}\ell^\a$. For models in  which this form of fuzziness admits phenomenological description in terms of distance fluctuations (which one would naturally expect, but needs to be checked in each specific model \cite{gacLRR}), one would then expect to find \cite{gacLRR,gw1} a strain noise $\s^2=\int\rmd\nu S^2(\nu)$ with spectral density $S(\nu) \propto c^\alpha (\ell_*)^{1-\alpha}$ $\times\nu^{-\frac{1+2\alpha}{2}}$ ($\nu$ here denoting the frequency), and this form of strain noise can be meaningfully constrained, even for very small $\ell_*$, using modern gravity-wave interferometers, such as LIGO and VIRGO \cite{gw1,gw2,ligovirgo}. Since $\alpha = d_{\rm S,H}^{\rm UV} / D$ (see above), we find for the UV dimension $d_{\rm S,H}^{\rm UV} \propto D \, \log(S\sqrt{\nu}/\ell_*)/\log(c/\nu\ell_*)$, and for a first order-of-magnitude estimate we can take as reference the LIGO sensitivity level of $S \sim 10^{-20}\, $m Hz$^{-1/2}$ at  $\nu \sim 10^3$ Hz. This allows to establish meaningful constraints even for ``Planckian values" of $\ell_*$: for example for  $\ell_* \simeq \lp$
 at  $10^3$ Hz one would expect fuzziness noise at the level of $10^{-20}\, $m Hz$^{-1/2}$ for $d_{\rm S,H}^{\rm UV} \sim 1.7$. So this is a rare case for quantum-gravity research where experimental sensitivities are at a level comparable to where we are with theoretical understanding, since most arguments point to $1.5 \lesssim \ds^{\rm UV} \lesssim 2.5$.

\section*{Acknowledgments} 

G.C.\ is under a Ram\'on y Cajal contract and is supported by the I+D grant FIS2014-54800-C2-2-P. M.R.\ thanks Instituto de Estructura de la Materia (CSIC) for the hospitality during the first stage of elaboration of this work. The contribution of G.C.\ and M.R.\ is based upon work from COST Action CA15117, supported by COST (European Cooperation in Science and Technology).

\end{document}